# Maximum Pseudolikelihood Estimation for Model-Based Clustering of Time Series Data

Hien D. Nguyen[12*], Geoffrey J. McLachlan[1], Pierre Orban[3],

Pierre Bellec[3], and Andrew L. Janke[2]

October 18, 2016

[1]School of Mathematics and Physics, University of Queensland. [2]Centre for Advanced Imaging, University of Queensland. [3]Centre de Recherche, Institut Universitaire de Geriatrie de Montreal. [*]Email: h.nguyen7@uq.edu.au.

**Abstract**

Mixture of autoregressions (MoAR) models provide a model-based approach to the clustering of time series data. The maximum likelihood (ML) estimation of MoAR models requires the evaluation of products of large numbers of densities of normal random variables. In practical scenarios, these products converge to zero as the length of the time series increases, and thus the ML estimation of MoAR models becomes infeasible without the use of numerical tricks. We propose a maximum pseudolikelihood (MPL) estimation approach as an alternative to the use of numerical tricks. The MPL estimator is proved to be consistent and can be computed via an EM (expectation–maximization) algorithm. Simulations are used to assess the performance of the MPL estimator against that of the ML estimator in cases where the latter was able to be calculated. An application to the clustering of time series data arising from a resting-state fMRI experiment is presented as a demonstration of the methodology.

# 1 Introduction

The simultaneous acquisition of large numbers of time series arises in many areas of modern science. This is especially true in the areas of biological and medical image analyses, where multiple time series are commonly acquired in electrocardiogram (ECG), electroencephalography (EEG), and functional magnetic resonance imaging (fMRI) experiments. In such experiments, hundreds to hundreds-of-thousands of time series can be acquired simultaneously, each often thousands of periods long. Upon acquisition, a common approach in such experiments is to organize the time series into similarity groups (clus-ters), based on their properties.

The clustering of time series data has received much attention in recent years. For example, the recent literature reports in Liao (2005) and Esling & Agon (2012) illustrate the breadth of research in the area.

It is clear from Esling & Agon (2012) that there are many potential directions for approaching the problem. Given the context of this article, we shall concentrate only on mixture-model based methods for clustering time series data. A brief review of recent developments in this direction is given below.

In Cadez et al. (2000), a mixture of Markov chains model was suggested for the clustering of data based on web browsing behavior, time-course gene expression, and red-blood cell cytograms. In Xiong & Yeung (2004) mixture of autoregressive moving-average regressions (MoARMA) models are suggested for the clustering of ECG, EEG, population, and temperature data. In Luan & Li (2003), Celeux et al. (2005), Ng et al. (2006), and Scharl et al. (2010), various specifications of mixture of mixed-effects models are suggested for the clustering of time-course gene expression data; Wang et al. (2012) extended the methodology of Ng et al. (2006) by considering moving-average errors. Lastly, Samé et al. (2011) suggested the use of mixture of linear experts for the clustering of electrical power consumption data.

Recently, Nguyen et al. (2016) have reconsidered the work of Xiong & Yeung (2004) and have proposed a mixture of autoregressions (MoAR) model for the clustering of spatially dependent time series data that arise from imaging based experiments. In their work, an MM algorithm [minorization–maximization; see Lange (2013, Ch. 8)] was proposed, which both monotonically increased the marginal likelihood objective function and which lead to global convergence to a stationary point of the log-marginal likelihood function. Furthermore, it was established that the maximum marginal likelihood estimator for the MoAR model was consistent under some regularity assumptions on the dependency structure of data. We note that "marginal likelihood" can be replaced by "likelihood" when the data are assumed to be independent.

The method presented in Nguyen et al. (2016) requires the evaluation of products of the form

$$\Pi\left(\boldsymbol{x}; \mu, \sigma^2\right) = \prod_{t=1}^{m} \phi\left(x_t; \mu, \sigma^2\right), \qquad (1)$$

where $\boldsymbol{x} = (x_1, ..., x_m)^T$ is a vector containing $m$ realizations of $X_t$, where $X_t$ arises from a finite normal mixture model with $g$ components (see McLachlan & Peel (2000, Ch. 3) regarding normal mixture models), and $\phi(x; \mu, \sigma^2)$ is a normal density function with mean $\mu \in \mathbb{R}$ and variance $\sigma^2 > 0$. Here the superscript $T$ indicates matrix transposition.

In standard application conditions, such products can decrease rapidly to values that are below usual machine precision for relatively small $m$, where $m$ is the length of the time series under analysis. Numerical tricks can be applied [e.g. Press et al. (2007, Sec. 16.1)] to avoid numerical underflows.

We present an alternative to these tricks via the use of pseudolikelihood (PL) functions.

In this article, we formulate the maximum pseudolikelihood (MPL) estimator of the MoAR model for long time series, under the MPL estimation framework of Arnold & Strauss (1991); see also Molenberghs & Verbeke (2005, Ch. 9). We prove that the MPL estimator is consistent under mild regularity conditions. Also, we construct an EM algorithm [expectation–maximization; Dempster et al. (1977)] for the MPL estimation of the MoAR model. We show that the algorithm monotonically increases the PL value at each iteration and consequently leads to global convergence to a stationary point of the log PL function.

Besides our algorithm and theoretical results, we also demonstrate the performance of our methodology via a simulation study. In this study, we demonstrate that the MPL estimator exhibits convergence towards the population parameter in finite samples. Also, we demonstrate that the MPL estimator can exhibit super-efficiency for the estimation of the mixing proportions, when compared to the maximum likelihood (ML) estimator. We further demonstrate our methodology via an application to clustering data arising from an fMRI experiment.

The remainder of the article proceeds as follows. In Section 2, we present the MoAR model, review the work of Nguyen et al. (2016), and examine the problems associated with the calculation of (1). In Section 3, we present the MPL estimator and construct an EM algorithm for its computation. In Section 4, we examine aspects of statistical inference that arise from the use of the MPL estimator. In Section 5, we present the results of our numerical simulations. In Section 6, we present an example analysis of an fMRI data set. Finally, conclusions are drawn in Section 7.

## 2 ML Estimation of MoAR Models

### 2.1 Mixture of Autoregressive Models

Let $\boldsymbol{Y}_s = (Y_{s1}, ..., Y_{sm})^T$ be a random vector of length $m$, indexed by $s = 1, ..., n$. Suppose that $Z_s \in \{1, ..., g\}$ is a latent random variable, such that $\mathbb{P}(Z_s = i) = \pi_i$, for $i = 1, ..., g$, where $\pi_i > 0$ and $\sum_{i=1}^{g} \pi_i = 1$. We say that $\boldsymbol{Y}_s$ arises from a $g$-component MoAR model of order $p$, if it can be characterized by the conditional density function

$$f\left(y_{st} | \boldsymbol{y}_{s(t)}, Z_s = i; \boldsymbol{\theta}\right) = \phi\left(y_{st}; \boldsymbol{\beta}_i^T \boldsymbol{y}_{s(t)}, \sigma_i^2\right), \quad (2)$$

where $\boldsymbol{y}_{s(t)} = (1, y_{s,t-1}, ..., y_{s,t-p})^T$, $\boldsymbol{\beta}_i = (\beta_{i0}, ..., \beta_{ip})^T \in \mathbb{R}^{p+1}$, and $\sigma_i^2 > 0$. Here $\boldsymbol{\theta} = \left(\pi_1, ..., \pi_{g-1}, \boldsymbol{\beta}_1^T, ..., \boldsymbol{\beta}_g^T, \sigma_1^2, ..., \sigma_g^2\right)^T$ is the model parameter vector.

Under the characterization (2), if we suppose that the first $p$ elements of $\boldsymbol{Y}_s$ are non-stochastic (i.e. $Y_{st} = y_{st}$, for $t = 1, ..., p$, almost everywhere), then we can further characterize $\boldsymbol{Y}_s$ via the joint conditional density function

$$f\left(\boldsymbol{y}_s|Z_s=i;\boldsymbol{\theta}\right)=\prod_{t=p+1}^{m}\phi\left(y_{st};\boldsymbol{\beta}_i^T\boldsymbol{y}_{s(t)},\sigma_i^2\right),$$

and hence the marginal density function

$$f\left(\boldsymbol{y}_s;\boldsymbol{\theta}\right)=\sum_{i=1}^{g}\pi_i\prod_{t=p+1}^{m}\phi\left(y_{st};\boldsymbol{\beta}_i^T\boldsymbol{y}_{s(t)},\sigma_i^2\right). \quad (3)$$

Using the characterization (3), we can write the likelihood and log likelihood of an IID (independent and identically distributed) sample $\boldsymbol{Y}_1,...,\boldsymbol{Y}_n$ as

$$\mathcal{L}_n\left(\boldsymbol{\theta}\right)=\prod_{s=1}^{n}f\left(\boldsymbol{y}_s;\boldsymbol{\theta}\right)$$

and

$$\ell_n\left(\boldsymbol{\theta}\right)=\sum_{s=1}^{n}\log\sum_{i=1}^{g}\pi_i\prod_{t=p+1}^{m}\phi\left(y_{st};\boldsymbol{\beta}_i^T\boldsymbol{y}_{s(t)},\sigma_i^2\right), \quad (4)$$

respectively.

Let the ML estimator $\hat{\boldsymbol{\theta}}_n$ be defined as an appropriate local-maximizer of (4). Due to the log-summation form of (4), it is not possible to deduce a closed form expression for $\hat{\boldsymbol{\theta}}_n$. As such, an iterative algorithm is required for the computation of $\hat{\boldsymbol{\theta}}_n$.

## 2.2 EM Algorithm for ML Estimation

Let $\boldsymbol{\theta}^{(0)}$ be the initial value of $\boldsymbol{\theta}$ for the application of the algorithm and let $\boldsymbol{\theta}^{(r)}$ be the $r$th iterate. In Nguyen et al. (2016), the following EM algorithm was considered for computation of $\hat{\boldsymbol{\theta}}_n$. We note that instead of using an EM algorithm, we could use for this problem a MM algorithm as, for example, in Nguyen et al. (2016) for their problem.

At the $(r+1)$th iteration, the updates are given by

$$\pi_i^{(r+1)} = n^{-1} \sum_{s=1}^{n} \tau_{is}\left(\boldsymbol{\theta}^{(r)}\right), \tag{5}$$

$$\boldsymbol{\beta}_i^{(r+1)} = \left[\sum_{s=1}^{n} \tau_{is}\left(\boldsymbol{\theta}^{(r)}\right) \sum_{t=p+1}^{m} \boldsymbol{y}_{s(t)}\boldsymbol{y}_{s(t)}^T\right]^{-1} \left[\sum_{s=1}^{n} \tau_{is}\left(\boldsymbol{\theta}^{(r)}\right) \sum_{t=p+1}^{m} \boldsymbol{y}_{s(t)}y_{st}\right], \tag{6}$$

and

$$\sigma_i^{2(r+1)} = \frac{\sum_{s=1}^{n} \tau_{is}\left(\boldsymbol{\theta}^{(r)}\right) \sum_{t=p+1}^{m} \left(y_{st} - \boldsymbol{y}_{s(t)}^T\boldsymbol{\beta}_i^{(r+1)}\right)^2}{(m-p)\sum_{s=1}^{n} \tau_{is}\left(\boldsymbol{\theta}^{(r)}\right)}, \tag{7}$$

for each $i = 1, ..., g$, where $\tau_{is}\left(\boldsymbol{\theta}\right) = \pi_i f\left(\boldsymbol{y}_s|Z_s=i;\boldsymbol{\theta}\right)/f\left(\boldsymbol{y}_s;\boldsymbol{\theta}\right)$ for each $s = 1, ..., n$.

As updates (5)–(7) are specified by an EM algorithm, the likelihood value increases monotonically at each iteration. Unfortunately, each iteration of the algorithm requires the computation of $\tau_{is}\left(\boldsymbol{\theta}^{(r)}\right)$ for every $s$, which requires the evaluation of multiple products of form (1). This can cause numerical underflow problems for large $m$ without the application of numerical tricks as mentioned earlier.

## 2.3 The Product Problem

We now consider the problem of computing (1) in a general context. Let $\boldsymbol{X} = (X_1, ..., X_m)^T$ be a vector of IID random variables with density function $f(x) = \sum_{i=1}^{g} \pi_i \phi(x; \mu_i, \sigma_i^2)$, where $\mu_i \in \mathbb{R}$, $\sigma_i^2 > 0$, $\pi_i^2 > 0$ for each $i$, and $\sum_{i=1}^{g} \pi_i = 1$. By independence and integration, we have the following result regarding the expectation of (1).

**Proposition 1.** *The expectation of (1) can be written as*

$$\mathbb{E}\left[\Pi\left(\boldsymbol{x}; \mu, \sigma^2\right)\right] = \left[\sum_{i=1}^{g} \pi_i \phi\left(\mu; \mu_i, \sigma^2 + \sigma_i^2\right)\right]^m. \tag{8}$$

Note that the summation in (8) is less than 1 if $\phi(\mu; \mu_i, \sigma^2 + \sigma_i^2) < 1$ for each $i$. For fixed $\sigma + \sigma_i^2$, $\phi(\mu; \mu_i, \sigma^2 + \sigma_i^2)$ attains a global maximum at $\mu = \mu_i = 0$; thus, the condition is fulfilled if we set $\sigma^2 + \sigma_i^2 > 1/(2\pi)$ for each $i$ (or simply $\sigma_i^2 > 1/(2\pi)$ since $\sigma^2 > 0$). Under such a condition, it is easy to see that (8) goes to zero as $m$ increases. This degeneration can be very rapid for models with high variances in each component. For example, consider the following result.

**Proposition 2.** *For any $\mu$ and $\sigma^2$, if $\underline{\sigma}^2 = \min_{i=1,...,g} \sigma_i^2$ and $\underline{\sigma}^2 > 1/(2\pi)$, then*

$$\mathbb{E}\left[\Pi\left(\boldsymbol{x}; \mu, \sigma^2\right)\right] \leq \left(2\pi\underline{\sigma}^2\right)^{-m/2}. \tag{9}$$

Thus, the numerical underflow can occur without the use of numerical tricks in a direct implementation of ML estimation via the EM algorithm.

We now consider an alternative to ML estimation that addresses the product problem without the use of numerical tricks.

## 3 Maximum Pseudolikelihood Estimation

### 3.1 Pseudolikelihood Function

Using characterization (2) and the PL definition of Arnold & Strauss (1991), we can write a PL function for a single time series as

$$g(\boldsymbol{y}_s; \boldsymbol{\theta}) = \prod_{t=p+1}^{m} \sum_{i=1}^{g} \pi_i \phi\left(y_{st}; \boldsymbol{\beta}_i^T \boldsymbol{y}_{s(t)}, \sigma_i^2\right) \tag{10}$$

for each $s$. We say "a" above since (10) is one of many possible PL functions that can be deduced from characterization (2). The chosen form of the PL function implicitly assumes that each of the $m - p$ random elements of $\boldsymbol{Y}_s$ can independently belong to each of the $g$ mixture components, conditioned on the $p$ previous elements. The specification allows for the construction of a log PL function

$$\begin{aligned}\mathcal{P}_n(\boldsymbol{\theta}) &= \sum_{s=1}^{n} \log g(\boldsymbol{y}_s; \boldsymbol{\theta}) \\ &= \sum_{s=1}^{n} \sum_{t=p+1}^{m} \log \sum_{i=1}^{g} \pi_i \phi\left(y_{st}; \boldsymbol{\beta}_i^T \boldsymbol{y}_{s(t)}, \sigma_i^2\right).\end{aligned} \tag{11}$$

Let the MPL estimator $\tilde{\boldsymbol{\theta}}_n$ be defined as an appropriate local maximum of (11). Like (4), (11) also contains terms of the log-summation form, and thus it is not possible to deduce a closed form expression of $\tilde{\boldsymbol{\theta}}_n$. We now present an EM algorithm for the iterative computation of the MPL estimate.

## 3.2 EM Algorithm for MPL Estimation

We can specify a so-called complete-data version of the PL function in that it can be viewed as a joint density of the observed time series data and their unobservable component-indicator variables that implies the PL function. The logarithm of this joint density (the complete-data log PL function) is given by

$$
\begin{aligned}
\mathcal{P}_n^c(\boldsymbol{\theta}) &= \sum_{i=1}^{g}\sum_{s=1}^{n}\sum_{t=p+1}^{m} \mathbb{I}(Z_{st}=i)\left[\log\pi_i + \log\phi\left(y_{st}; \boldsymbol{y}_{s(t)}^T\boldsymbol{\beta}_i, \sigma_i^2\right)\right], \\
&= \sum_{i=1}^{g}\log\pi_i \sum_{s=1}^{n}\sum_{t=p+1}^{m}\mathbb{I}(Z_{st}=i) \hspace{2em} (12)\\
&\quad -\frac{1}{2}\sum_{i=1}^{g}\log\sigma_i^2 \sum_{s=1}^{n}\sum_{t=p+1}^{m}\mathbb{I}(Z_{st}=i) \\
&\quad -\frac{1}{2}\sum_{i=1}^{g}\frac{1}{\sigma_i^2}\sum_{s=1}^{n}\sum_{t=p+1}^{m}\mathbb{I}(Z_{st}=i)\left(y_{st} - y_{s(t)}^T\boldsymbol{\beta}_i\right)^2 + \mathcal{C},
\end{aligned}
$$

where $\mathcal{C}$ gathers up constants that do not depend on $\boldsymbol{\theta}$ and $Z_{st} \in \{1, ..., g\}$ is the component membership of time point $t$ of series $\boldsymbol{Y}_s$, given the previous $p$ terms. Here, $\mathbb{I}(A)$ is the indicator variable that takes value 1 if proposition $A$ is true and 0 otherwise.

Starting from some initial value $\boldsymbol{\theta}^{(0)}$, the expectation of (12), computed using $\boldsymbol{\theta}^{(r)}$ for $\boldsymbol{\theta}$, can be written as

$$
\begin{aligned}
\mathcal{Q}\left(\boldsymbol{\theta}; \boldsymbol{\theta}^{(r)}\right) &= \sum_{i=1}^{g}\sum_{s=1}^{n}\sum_{t=p+1}^{m} \tau_{ist}\left(\boldsymbol{\theta}^{(r)}\right)\left[\log \pi_i + \log \phi\left(y_{st}; \boldsymbol{y}_{s(t)}^T \boldsymbol{\beta}_i, \sigma_i^2\right)\right], \\
&= \sum_{i=1}^{g} \log \pi_i \sum_{s=1}^{n}\sum_{t=p+1}^{m} \tau_{ist}\left(\boldsymbol{\theta}^{(r)}\right) \\
&\quad -\frac{1}{2}\sum_{i=1}^{g} \log \sigma_i^2 \sum_{s=1}^{n}\sum_{t=p+1}^{m} \tau_{ist}\left(\boldsymbol{\theta}^{(r)}\right) \\
&\quad -\frac{1}{2}\sum_{i=1}^{g}\frac{1}{\sigma_i^2}\sum_{s=1}^{n}\sum_{t=p+1}^{m} \tau_{ist}\left(\boldsymbol{\theta}^{(r)}\right)\left(y_{st} - \boldsymbol{y}_{s(t)}^T \boldsymbol{\beta}_i\right)^2 + \mathcal{C},
\end{aligned}
\tag{13}
$$

where

$$
\tau_{ist}\left(\boldsymbol{\theta}\right) = \frac{\pi_i \phi\left(y_{st}; \boldsymbol{\beta}_i^T \boldsymbol{y}_{s(t)}, \sigma_i^2\right)}{\sum_{j=1}^{g} \pi_j \phi\left(y_{st}; \boldsymbol{\beta}_j^T \boldsymbol{y}_{s(t)}, \sigma_j^2\right)}.
\tag{14}
$$

The posterior probability is the conditional probability that $y_{st}$ belongs to the $i$th component given $y_{st}$ and $\boldsymbol{y}_{s(t)}$ for $i = 1, ..., g$; $s = 1, ..., n$; and $t = p+1, ..., m$.

To perform the M-step, we maximize (13) under the restriction $\sum_{i=1}^{g} \pi_i = 1$, by constructing the Lagrangian $\Lambda\left(\boldsymbol{\theta}, \lambda\right) = \mathcal{U}\left(\boldsymbol{\theta}; \boldsymbol{\theta}^{(r)}\right) + \lambda\left(\sum_{i=1}^{g} \pi_i - 1\right)$ and solving the equation corresponding to the first-order condition $\nabla \Lambda = \boldsymbol{0}$, where $\nabla$ is the gradient operator. This yields the updates

$$
\pi_i^{(r+1)} = n^{-1} \sum_{s=1}^{n}\sum_{t=p+1}^{m} \tau_{ist}\left(\boldsymbol{\theta}^{(r)}\right),
\tag{15}
$$

$$\boldsymbol{\beta}_i^{(r+1)} = \left[ \sum_{s=1}^{n} \sum_{t=p+1}^{m} \tau_{ist}\left(\boldsymbol{\theta}^{(r)}\right) \boldsymbol{y}_{s(t)} \boldsymbol{y}_{s(t)}^T \right]^{-1} \left[ \sum_{s=1}^{n} \sum_{t=p+1}^{m} \tau_{ist}\left(\boldsymbol{\theta}^{(r)}\right) \boldsymbol{y}_{s(t)} y_t \right], \quad (16)$$

and

$$\sigma_i^{2(r+1)} = \frac{\sum_{s=1}^{n} \sum_{t=p+1}^{m} \tau_{ist}\left(\boldsymbol{\theta}^{(r)}\right) \left(y_{st} - \boldsymbol{y}_{s(t)}^T \boldsymbol{\beta}_i^{(r+1)}\right)^2}{\sum_{s=1}^{n} \sum_{t=p+1}^{m} \tau_{ist}\left(\boldsymbol{\theta}^{(r)}\right)} \quad (17)$$

for each $i$. Closely following the proof of Nguyen & McLachlan (2015, Thm. 2), we obtain the following analogue to Nguyen et al. (2016, Prop. 3).

**Proposition 3.** *Given $\boldsymbol{\theta}^{(r)}$, if $\boldsymbol{\theta}^{(r+1)}$ is obtained via the updates (15)–(17) and*

$$\Theta = \left\{ \pi_1, ..., \pi_{g-1} : \sum_{i=1}^{g-1} \pi_i < 1, \pi_i > 0 \right\} \times \mathbb{R}^{g(p+1)} \times (0, \infty)^g, \quad (18)$$

*then*

$$\boldsymbol{\theta}^{(r+1)} = \arg\max_{\boldsymbol{\theta} \in \Theta} \mathcal{Q}\left(\boldsymbol{\theta}; \boldsymbol{\theta}^{(r)}\right).$$

Proposition 3 implies that the log PL function monotonically increases at each iteration when the update steps (15)–(17) are used.

## 3.3 Global Convergence via the EM Algorithm

Given some initial value $\boldsymbol{\theta}^{(0)}$, the EM algorithm defined by updates (15)–(17) is run for some fixed number of iterations or until some convergence criterion is met, whereupon the final iterate of the algorithm is declared the MPL estimate $\tilde{\boldsymbol{\theta}}_n$; see Lange (2013, Sec. 11.5) for a description of various stoping criteria and their relative merits.

Let $\boldsymbol{\theta}^* = \lim_{r\to\infty} \boldsymbol{\theta}^{(r)}$ be a limit point of the EM algorithm, starting from some initial value $\boldsymbol{\theta}^{(0)}$. It is known that the EM algorithm is a special case of the MM algorithm [cf. Razaviyayn et al. (2013, Sec. 8.5)]. As such, the following theorem regarding the limit points of the EM algorithm can be adapted from the MM algorithm theory of Razaviyayn et al. (2013).

**Theorem 1.** *Starting from some initial value $\boldsymbol{\theta}^{(0)}$, if $\boldsymbol{\theta}^*$ is a finite limit-point of the sequence $\boldsymbol{\theta}^{(r)}$, obtained via updates (15)–(17), then $\boldsymbol{\theta}^*$ is a saddle-point or local-maximum of (11).*

As with the log likelihood function from Nguyen et al. (2016), the log PL function is also unbounded. Because of this, the choice of initial value can be crucial to the success of the algorithm in finding an appropriate maximizer of (11). An example of a procedure that can be used to find good initial values is described in McLachlan & Peel (2000, Sec. 2.12.2).

## 4 Statistical Inference

### 4.1 Consistency of MPL Estimator

Under usual regularity conditions, the MPL estimator is known to be consistent; see, for example, Arnold & Strauss (1991). Unfortunately, the log PL function is not identifiable and thus the usual asymptotic formulations cannot be used. As such, we apply Amemiya (1985, Thm. 4.1.2) to derive a result analogous to Nguyen et al. (2016, Thm. 2).

**Theorem 2.** *Let $Y_1, ..., Y_n$ be an IID random sample, such that for each $s$, $Y_s$ arises from a population with density function $f(y_s; \theta_0)$, where $\theta_0$ is a strict-local maximizer of $\mathbb{E} \log g(y_s; \theta_0)$. If $\Theta_n = \{\theta : \nabla \mathcal{P}_n = 0\}$ (where we take $\Theta_n = \{\bar{\theta}\}$, for some $\bar{\theta} \in \Theta$, if $\nabla \mathcal{P}_n = 0$ has no solution), then for any $\epsilon > 0$,*

$$\lim_{n \to \infty} \mathbb{P}\left[\inf_{\theta \in \Theta_n} (\theta - \theta_0)^T (\theta - \theta_0) > \epsilon\right] = 0.$$

We omit the proof of Theorem 2, as it follows closely the proof of Nguyen et al. (2016, Thm. 2). We make the following remarks regarding Theorem 2.

First, note that the theorem implies that the consistent roots of the log PL function are not necessarily the consistent roots of the log likelihood equation. In many situations the two sets of roots will correspond; Kenne Pagui et al. (2015) present a result regarding conditions under which such correspondence occurs. Unfortunately, it is difficult to verify the score and information conditions of Kenne Pagui et al. (2015), due to the nature of mixture-model densities. Second, the theorem only suggests that there may exist multiple roots of the log PL equation, of which one is consistent; as noted earlier, it is advisable to search for good initial values that lead to the correct root. Finally, the theorem can be extended to dependent identically distributed samples via conditioning on the dependence structure of $Y_1, ..., Y_n$. For example, like in Nguyen et al. (2016, Thm. 2), one can assume ergodicity or strong-mixing conditions.

## 4.2 Cluster Analysis

When performing model-based clustering, one would generally utilize the plugin Bayes' rule for risk-minimal allocation. Let $\tilde{z}_{sn} \in \{1,..,g\}$ be the plugin Bayes' allocation of observation $s$ and note that $\tau_{is}\left(\tilde{\boldsymbol{\theta}}_n\right)$ is the estimated posterior probability of observation $s$ belonging to cluster $i$. In the current context, observation $\boldsymbol{Y}_s$ can be allocated via the plugin Bayes' rule

$$\tilde{z}_{sn} = \arg\max_{i=1,...,g} \tau_{is}\left(\tilde{\boldsymbol{\theta}}_n\right). \tag{19}$$

We cannot guarantee the convergence of (19) to the same allocation of $\boldsymbol{Y}_s$ as that obtained via the ML estimator, since we cannot establish the equivalence between $\tilde{\boldsymbol{\theta}}_n$ and $\hat{\boldsymbol{\theta}}_n$. Furthermore, the computation of (19) requires products of form (1), which we are trying to avoid.

Unfortunately, we cannot overcome the first of these two caveats in a simple manner. Fortunately, the second can be addressed via an approximation. Using (14), we say that $\bar{z}_{sn} \in \{1,...,g\}$ is the pseudoallocation of $\boldsymbol{Y}_s$ and define it as

$$\bar{z}_{sn} = \arg\max_{i=1,...,g} \bar{\tau}_{is}\left(\tilde{\boldsymbol{\theta}}_n\right), \tag{20}$$

where $\bar{\tau}_{is}\left(\tilde{\boldsymbol{\theta}}_n\right) = (m-p)^{-1}\sum_{t=p+1}^{m} \tau_{ist}\left(\tilde{\boldsymbol{\theta}}_n\right)$ and $\bar{z}_{sn}$ is the cluster pseudoallocation of $\boldsymbol{Y}_s$.

Define the strong-mixing rate of observation $\boldsymbol{Y}_s$ over time as

$$\alpha_s(k) = \sup_m \left\{ |\mathbb{P}(A \cap B) - \mathbb{P}(A)\mathbb{P}(B)| : A \in \mathcal{F}_1^m, B \in \mathcal{F}_{m+k}^\infty \right\},$$

where $\mathcal{F}_a^b$ is the $\sigma$-algebra generated by $Y_{sa}, ..., Y_{sb}$, for $a < b$. The following result establishes the $\alpha$-mixing of $\boldsymbol{Y}_s$, for each $s$, and thus the convergence of $\bar{\tau}_{is}\left(\tilde{\boldsymbol{\theta}}_n\right)$ to nontrivial limits under reasonable assumptions on the conditional characterizations (2).

**Proposition 4.** *If the characteristic polynomials $\zeta^p - \sum_{k=1}^p \beta_{ik}\zeta^{p-k} = 0$ have roots inside the unit circle (with respect to $\zeta$), for each $i = 1, ..., g$, then for each $s = 1, ..., n$;*

**(a)** *the time series $\boldsymbol{Y}_s$ is strong-mixing.*

**(b)** *the cluster pseudoallocation $\bar{\tau}_{is}\left(\tilde{\boldsymbol{\theta}}_n\right)$ converges to $\mathbb{E}\left[\tau_{ist}\left(\tilde{\boldsymbol{\theta}}_n\right)\right]$, for each $i = 1, ..., g$.*

*Proof.* The hypothesis of the proposition guarantees that each of the conditional characterizations (2) are $\alpha$-mixing [cf. Athreya & Pantula (1986)]. Thus, if we denote the strong-mixing rate of $\boldsymbol{Y}_s$ conditioned on $Z_s = i$ as $\alpha_{is}(k)$, then $\alpha_{is}(k) \to 0$ as $m \to \infty$ for each $i$. Since $\boldsymbol{Y}_s$ can only exhibit one of the $g$ different behaviors of the conditional characterizations, we have

$$\alpha_s(k) \leq \max \alpha_{is}(k) \leq \sum_{i=1}^g \alpha_{is}(k) \to 0$$

as $m \to 0$; this implies part (a).

Table 1: Parameter vectors of C1–C4, as used in S1 and S2.

| Class | $\beta_{i0}$ | $\beta_{i1}$ | $\beta_{i2}$ | $\sigma_i^2$ |
|:---:|:---:|:---:|:---:|:---:|
| $C_1$ | 0 | 0 | 0.25 | 1 |
| $C_2$ | 0 | 0 | -0.25 | 1 |
| $C_3$ | 0 | 0.25 | 0 | 1 |
| $C_4$ | 0 | -0.25 | 0 | 1 |

Since $\tau_{ist}\left(\tilde{\boldsymbol{\theta}}_n\right)$ are continuous functions of finitely many terms of $\boldsymbol{Y}_s$, $\tau_{ist}\left(\tilde{\boldsymbol{\theta}}_n\right)$ are also strong-mixing [cf. White (2001, Thm. 3.49)] for each $i$ and $s$. Because $\tau_{ist}\left(\tilde{\boldsymbol{\theta}}_n\right)$ is bounded, it also has all of its moments, and thus White (2001, Corr. 3.48) establishes the convergence of $\bar{\tau}_{is}\left(\tilde{\boldsymbol{\theta}}_n\right)$ to $\mathbb{E}\left[\tau_{ist}\left(\tilde{\boldsymbol{\theta}}_n\right)\right]$, as $m \to \infty$. This proves part (b). □

In general, we do not expect the limit $\mathbb{E}\left[\tau_{ist}\left(\tilde{\boldsymbol{\theta}}_n\right)\right]$ to equate to $\mathbb{E}\left[\tau_{is}\left(\tilde{\boldsymbol{\theta}}_n\right)\right]$. We compare the performances of rules (19) and (20) in the next section.

# 5 Numerical Simulations

## 5.1 Simulation Setup

We report on two numerical simulation studies designated S1 and S2. In both studies, refer to the classes of generative models $(C_1–C_4)$ that are reported in Table 1. Examples of series of length 100 from each class are plotted in Figure 1.

In S1, we generate $n$ time series of length $m$ from classes $C_1$ and $C_2$ with probabilities $\pi_1 = \pi_2 = 0.5$, where $m, n = 100, 200, 500, 1000$. This is repeated $N = 100$ times for each combination of $m$ and $n$. In S2, we generate $n$ time series of length $m$, for the same range of $m$ and $n$ as in S1, from classes $C_1$–$C_4$ with probabilities $\pi_i = 0.25$, for each $i = 1, .., 4$. This is also repeated $N = 100$ times for each $m$ and $n$.

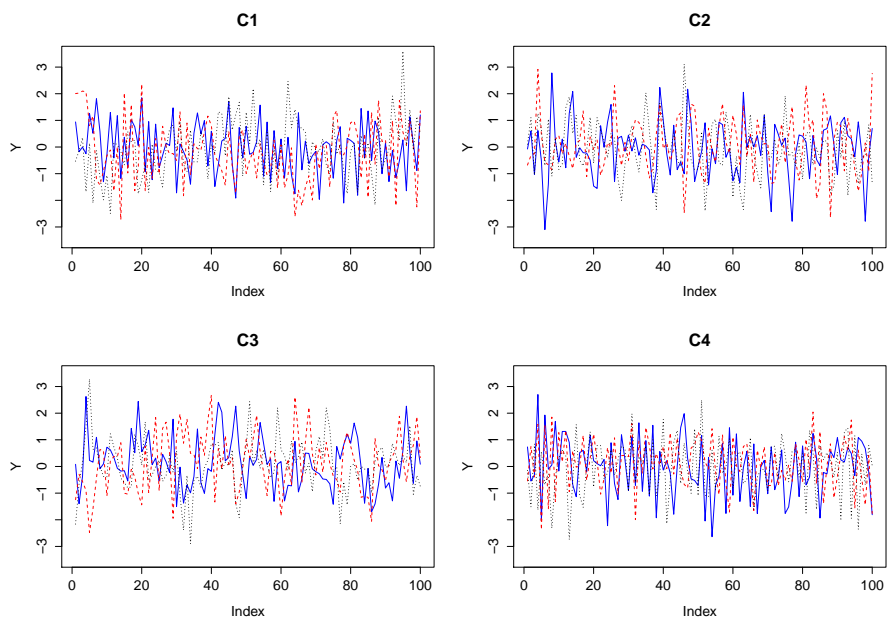

Figure 1: Three realizations of time series of length $m = 100$ from each of the classes C1–C4, as described in Table 1.

For each combination and each study, we compute the MPL estimate and calculate the mean squared error (MSE) $N^{-1} \sum_{j=1}^{N} \left( \tilde{\theta}_{nk} - \theta_{0k} \right)^2$ for each parameter element, where $\tilde{\theta}_{nk}$ and $\theta_{0k}$ denote the $k$th element of the MPL estimate $\tilde{\boldsymbol{\theta}}_n$ and the true parameter vector $\boldsymbol{\theta}_0$ (as given in Table 1), respectively. Here $k = 1, ..., 10$ in S1 and $k = 1, ..., 20$ in S2. The MSE results for S1 and S2 are presented in Table 2, and Tables 3 and 4, respectively.

Further, we also measure the similarity of the pseudoallocation (20) in comparison to the cluster allocation (19). We make comparisons via the average similarity measurement $(nN)^{-1} \sum_{j=1}^{N} \sum_{s=1}^{n} \mathbb{I}\left( \bar{c}_{sn} = \tilde{c}_{sn} \right)$, where $\mathbb{I}(A)$ is an indicator function that takes value 1 if proposition $A$ is true, and 0 otherwise. The results for both studies are presented in Table 5.

Finally, we assess the efficiency of the MPL estimator relative to the ML estimator. We do this by computing the ML estimate and calculating the MSE $N^{-1} \sum_{j=1}^{N} \left( \hat{\theta}_{nk} - \theta_{0k} \right)^2$ for each parameter element, where $\hat{\theta}_{nk}$ is the $k$th element of the ML estimate $\hat{\boldsymbol{\theta}}_n$. We then compute the ratio of the ML MSE to the MPL MSE. The results for S1 and S2 are reported in Table 6, and Tables 7 and 8, respectively.

All simulations are conducted in the $R$ statistical programming environment (R Core Team, 2013). The autoregressive time series are generated

Table 2: S1 MSE results for the MPL estimate vector versus the true parameter vector, averaged over $N = 100$ repetitions. Here, $aEb = a \times 10^b$.

| $m$ | $n$ | $\pi_1$ | $\pi_2$ | $\beta_{10}$ | $\beta_{11}$ | $\beta_{12}$ | $\beta_{20}$ | $\beta_{21}$ | $\beta_{22}$ | $\sigma_1^2$ | $\sigma_2^2$ |
|---|---|---|---|---|---|---|---|---|---|---|---|
| 100 | 100 | 1.4E-05 | 1.4E-05 | 3.5E-04 | 3.1E-04 | 2.9E-04 | 3.5E-04 | 3.1E-04 | 2.8E-04 | 6.9E-04 | 7.3E-04 |
| 100 | 200 | 6.5E-06 | 6.5E-06 | 1.6E-04 | 1.4E-04 | 1.3E-04 | 1.6E-04 | 1.4E-04 | 1.3E-04 | 3.2E-04 | 2.9E-04 |
| 100 | 500 | 4.0E-06 | 4.0E-06 | 4.6E-05 | 5.6E-05 | 5.5E-05 | 4.5E-05 | 5.6E-05 | 5.1E-05 | 1.4E-04 | 1.1E-04 |
| 100 | 1000 | 1.3E-06 | 1.3E-06 | 2.9E-05 | 3.1E-05 | 1.8E-05 | 2.8E-05 | 3.1E-05 | 2.0E-05 | 5.8E-05 | 6.5E-05 |
| 200 | 100 | 1.7E-05 | 1.7E-05 | 1.6E-04 | 1.5E-04 | 1.0E-04 | 1.6E-04 | 1.5E-04 | 1.0E-04 | 4.7E-04 | 3.5E-04 |
| 200 | 200 | 6.2E-06 | 6.2E-06 | 9.5E-05 | 7.4E-05 | 5.5E-05 | 9.5E-05 | 7.2E-05 | 5.6E-05 | 1.9E-04 | 1.9E-04 |
| 200 | 500 | 2.8E-06 | 2.8E-06 | 3.2E-05 | 3.5E-05 | 2.1E-05 | 3.1E-05 | 3.5E-05 | 2.2E-05 | 8.4E-05 | 1.0E-04 |
| 200 | 1000 | 1.5E-06 | 1.5E-06 | 1.6E-05 | 1.3E-05 | 1.2E-05 | 1.6E-05 | 1.3E-05 | 1.2E-05 | 4.2E-05 | 4.2E-05 |
| 500 | 100 | 1.1E-05 | 1.1E-05 | 5.5E-05 | 5.9E-05 | 4.7E-05 | 5.5E-05 | 5.9E-05 | 4.3E-05 | 1.6E-04 | 1.7E-04 |
| 500 | 200 | 6.8E-06 | 6.8E-06 | 3.5E-05 | 2.6E-05 | 2.2E-05 | 3.5E-05 | 2.5E-05 | 2.5E-05 | 8.5E-05 | 9.2E-05 |
| 500 | 500 | 3.0E-06 | 3.0E-06 | 1.2E-05 | 1.2E-05 | 9.7E-06 | 1.2E-05 | 1.2E-05 | 1.1E-05 | 3.5E-05 | 3.9E-05 |
| 500 | 1000 | 1.2E-06 | 1.2E-06 | 6.0E-06 | 4.5E-06 | 4.5E-06 | 6.0E-06 | 4.6E-06 | 4.7E-06 | 1.8E-05 | 1.8E-05 |
| 1000 | 100 | 1.3E-05 | 1.3E-05 | 3.0E-05 | 2.8E-05 | 3.0E-05 | 3.0E-05 | 2.8E-05 | 3.3E-05 | 7.5E-05 | 9.7E-05 |
| 1000 | 200 | 6.6E-06 | 6.6E-06 | 1.8E-05 | 1.4E-05 | 1.5E-05 | 1.7E-05 | 1.4E-05 | 1.4E-05 | 4.6E-05 | 4.8E-05 |
| 1000 | 500 | 2.8E-06 | 2.8E-06 | 6.9E-06 | 6.2E-06 | 5.6E-06 | 6.8E-06 | 6.0E-06 | 5.9E-06 | 2.6E-05 | 2.6E-05 |
| 1000 | 1000 | 1.0E-06 | 1.0E-06 | 2.9E-06 | 3.0E-06 | 2.2E-06 | 2.9E-06 | 2.9E-06 | 2.5E-06 | 1.1E-05 | 1.1E-05 |

Table 3: S2 MSE results for the first ten MPL estimate elements versus the respective true parameter elements, averaged over $N = 100$ repetitions. Here, $a\text{E}b = a \times 10^b$.

| $m$ | $n$ | $\pi_1$ | $\pi_2$ | $\pi_3$ | $\pi_4$ | $\beta_{10}$ | $\beta_{11}$ | $\beta_{12}$ | $\beta_{20}$ | $\beta_{21}$ | $\beta_{22}$ |
|---|---|---|---|---|---|---|---|---|---|---|---|
| 100 | 100 | 8.4E-06 | 7.8E-06 | 2.3E-06 | 2.5E-06 | 4.8E-04 | 3.6E-04 | 4.3E-04 | 4.6E-04 | 4.5E-04 | 5.9E-04 |
| 100 | 200 | 6.7E-06 | 6.1E-06 | 1.6E-06 | 1.2E-06 | 2.3E-04 | 1.4E-04 | 2.7E-04 | 2.0E-04 | 2.0E-04 | 3.2E-04 |
| 100 | 500 | 6.1E-06 | 4.9E-06 | 5.2E-07 | 5.3E-07 | 1.1E-04 | 6.2E-05 | 1.1E-04 | 1.1E-04 | 5.3E-05 | 1.5E-04 |
| 100 | 1000 | 6.0E-06 | 4.8E-06 | 2.7E-07 | 2.4E-07 | 6.1E-05 | 3.9E-05 | 6.3E-05 | 6.3E-05 | 4.0E-05 | 6.5E-05 |
| 200 | 100 | 8.5E-06 | 7.6E-06 | 2.3E-06 | 2.4E-06 | 2.3E-04 | 1.3E-04 | 3.2E-04 | 2.6E-04 | 2.0E-04 | 4.0E-04 |
| 200 | 200 | 6.0E-06 | 5.4E-06 | 9.0E-07 | 7.3E-07 | 1.4E-04 | 6.5E-05 | 2.0E-04 | 1.5E-04 | 1.2E-04 | 1.9E-04 |
| 200 | 500 | 5.5E-06 | 4.4E-06 | 5.2E-07 | 4.4E-07 | 5.1E-05 | 2.6E-05 | 7.0E-05 | 5.7E-05 | 3.8E-05 | 7.0E-05 |
| 200 | 1000 | 5.6E-06 | 4.4E-06 | 2.4E-07 | 2.4E-07 | 2.5E-05 | 1.9E-05 | 4.8E-05 | 3.0E-05 | 2.4E-05 | 4.2E-05 |
| 500 | 100 | 7.4E-06 | 6.3E-06 | 2.1E-06 | 1.9E-06 | 1.0E-04 | 6.2E-05 | 1.3E-04 | 9.2E-05 | 9.6E-05 | 1.9E-04 |
| 500 | 200 | 6.9E-06 | 5.0E-06 | 1.1E-06 | 1.0E-06 | 5.6E-05 | 3.2E-05 | 8.6E-05 | 6.9E-05 | 4.5E-05 | 1.0E-04 |
| 500 | 500 | 5.2E-06 | 4.4E-06 | 3.5E-07 | 3.0E-07 | 2.3E-05 | 1.0E-05 | 3.0E-05 | 2.3E-05 | 1.9E-05 | 3.8E-05 |
| 500 | 1000 | 5.6E-06 | 4.7E-06 | 1.9E-07 | 1.9E-07 | 1.0E-05 | 6.9E-06 | 2.0E-05 | 1.0E-05 | 1.0E-05 | 2.7E-05 |
| 1000 | 100 | 6.1E-06 | 5.6E-06 | 1.3E-06 | 1.4E-06 | 5.7E-05 | 4.1E-05 | 1.1E-04 | 5.4E-05 | 4.8E-05 | 1.3E-04 |
| 1000 | 200 | 5.8E-06 | 5.0E-06 | 7.6E-07 | 7.7E-07 | 3.0E-05 | 1.7E-05 | 4.9E-05 | 2.6E-05 | 2.9E-05 | 6.7E-05 |
| 1000 | 500 | 5.3E-06 | 4.4E-06 | 2.8E-07 | 2.4E-07 | 9.3E-06 | 7.6E-06 | 2.8E-05 | 1.1E-05 | 1.2E-05 | 3.5E-05 |
| 1000 | 1000 | 5.3E-06 | 4.4E-06 | 1.8E-07 | 1.9E-07 | 5.0E-06 | 2.7E-06 | 1.4E-05 | 5.4E-06 | 5.7E-06 | 1.3E-05 |

Table 4: S2 MSE results for the last ten MPL estimate elements versus the respective true parameter elements, averaged over $N = 100$ repetitions. Here, $a\text{E}b = a \times 10^b$.

| $m$ | $n$ | $\beta_{30}$ | $\beta_{31}$ | $\beta_{32}$ | $\beta_{40}$ | $\beta_{41}$ | $\beta_{42}$ | $\sigma_1^2$ | $\sigma_2^2$ | $\sigma_3^2$ | $\sigma_4^2$ |
|---|---|---|---|---|---|---|---|---|---|---|---|
| 100 | 100 | 5.5E-04 | 6.7E-04 | 4.6E-04 | 4.8E-04 | 8.1E-04 | 4.1E-04 | 1.7E-03 | 1.6E-03 | 1.6E-03 | 1.6E-03 |
| 100 | 200 | 3.2E-04 | 3.4E-04 | 1.6E-04 | 3.0E-04 | 3.6E-04 | 2.2E-04 | 7.1E-04 | 8.2E-04 | 7.8E-04 | 6.7E-04 |
| 100 | 500 | 9.4E-05 | 1.6E-04 | 7.6E-05 | 1.0E-04 | 1.5E-04 | 6.5E-05 | 3.7E-04 | 4.1E-04 | 2.8E-04 | 2.5E-04 |
| 100 | 1000 | 3.6E-05 | 8.7E-05 | 5.9E-05 | 4.7E-05 | 9.3E-05 | 4.3E-05 | 1.8E-04 | 2.5E-04 | 1.2E-04 | 1.1E-04 |
| 200 | 100 | 3.1E-04 | 3.2E-04 | 1.6E-04 | 3.1E-04 | 3.4E-04 | 1.8E-04 | 8.4E-04 | 8.4E-04 | 7.1E-04 | 5.7E-04 |
| 200 | 200 | 1.4E-04 | 1.9E-04 | 6.9E-05 | 1.6E-04 | 1.8E-04 | 1.3E-04 | 3.2E-04 | 5.2E-04 | 3.2E-04 | 3.7E-04 |
| 200 | 500 | 5.5E-05 | 1.0E-04 | 4.1E-05 | 5.8E-05 | 1.1E-04 | 3.7E-05 | 2.1E-04 | 2.5E-04 | 1.3E-04 | 1.2E-04 |
| 200 | 1000 | 2.7E-05 | 6.7E-05 | 3.0E-05 | 2.0E-05 | 6.1E-05 | 3.2E-05 | 1.7E-04 | 2.5E-04 | 6.2E-05 | 7.6E-05 |
| 500 | 100 | 8.6E-05 | 1.8E-04 | 9.2E-05 | 8.8E-05 | 2.0E-04 | 9.0E-05 | 3.4E-04 | 4.6E-04 | 3.5E-04 | 3.2E-04 |
| 500 | 200 | 5.7E-05 | 1.5E-04 | 5.3E-05 | 5.5E-05 | 1.6E-04 | 4.9E-05 | 2.2E-04 | 2.9E-04 | 1.2E-04 | 1.2E-04 |
| 500 | 500 | 2.0E-05 | 6.7E-05 | 2.4E-05 | 1.8E-05 | 6.8E-05 | 2.4E-05 | 1.4E-04 | 2.0E-04 | 6.0E-05 | 5.8E-05 |
| 500 | 1000 | 1.1E-05 | 4.2E-05 | 2.2E-05 | 1.1E-05 | 4.3E-05 | 1.8E-05 | 1.2E-04 | 1.7E-04 | 2.7E-05 | 3.0E-05 |
| 1000 | 100 | 5.1E-05 | 1.1E-04 | 4.9E-05 | 4.7E-05 | 1.1E-04 | 6.0E-05 | 2.0E-04 | 2.7E-04 | 9.9E-05 | 9.4E-05 |
| 1000 | 200 | 2.1E-05 | 8.6E-05 | 3.1E-05 | 2.4E-05 | 9.0E-05 | 3.6E-05 | 1.3E-04 | 2.0E-04 | 6.5E-05 | 6.9E-05 |
| 1000 | 500 | 9.4E-06 | 5.1E-05 | 1.9E-05 | 8.7E-06 | 5.1E-05 | 1.7E-05 | 1.2E-04 | 1.8E-04 | 2.7E-05 | 2.7E-05 |
| 1000 | 1000 | 4.3E-06 | 3.8E-05 | 1.4E-05 | 4.5E-06 | 3.8E-05 | 1.5E-05 | 1.0E-04 | 1.6E-04 | 1.6E-05 | 1.5E-05 |

Table 5: Similarity measurements of the pseudoallocation (20) versus the cluster allocation (19) in S1 and S2.

| S1 | | $m$ | | | |
|---|---|---|---|---|---|
| | | 100 | 200 | 500 | 1000 |
| | 100 | 0.9929 | 0.9994 | 1.0000 | 1.0000 |
| $n$ | 200 | 0.9947 | 0.9996 | 1.0000 | 1.0000 |
| | 500 | 0.9963 | 0.9998 | 1.0000 | 1.0000 |
| | 1000 | 0.9973 | 0.9999 | 1.0000 | 1.0000 |
| S2 | | $m$ | | | |
| | | 100 | 200 | 500 | 1000 |
| | 100 | 0.8958 | 0.9667 | 0.9983 | 1.0000 |
| $n$ | 200 | 0.9238 | 0.9758 | 0.9983 | 1.0000 |
| | 500 | 0.9336 | 0.9788 | 0.9995 | 1.0000 |
| | 1000 | 0.9382 | 0.9788 | 0.9995 | 1.0000 |

using the *arima.sim* function in $R$. The EM algorithms are programmed in $R$, with the log PL value evaluations and EM algorithm updates coded in $C$ via the *Rcpp* and *RcppArmadillo* packages (Eddelbuettel, 2013).

## 5.2 Results

Upon inspection of Tables 2–4, we observe that there there is a general decreasing trend in terms of both increases in $m$ and $n$, in all parameter elements. We see that the decreasing trend in $m$ is more gradual than in $n$. Furthermore, the MSEs of the mixing proportions (i.e. $\pi_i$) appear to be not affected by the changes in $m$. Also, we see that the decrease of the MSE with respect to $n$ is predicted by Theorem 2.

The results from Table 5 indicate that the similarity of pseudoallocations and the cluster allocations increase with $m$. Here, we observe that in S1,

Table 6: Relative efficiency in S1, as measured by the ratio of the ML MSE to the MPL MSE for each parameter element, computed over $N = 100$ repetitions. Here, $a\mathrm{E}b = a \times 10^b$.

| $m$ | $n$ | $\pi_1$ | $\pi_2$ | $\beta_{10}$ | $\beta_{11}$ | $\beta_{12}$ | $\beta_{20}$ | $\beta_{21}$ | $\beta_{22}$ | $\sigma_1^2$ | $\sigma_2^2$ |
|---|---|---|---|---|---|---|---|---|---|---|---|
| 100 | 100 | 1.9E+02 | 1.9E+02 | 7.0E-01 | 5.8E-01 | 8.7E-01 | 5.7E-01 | 5.4E-01 | 7.1E-01 | 6.5E-01 | 5.1E-01 |
| 100 | 200 | 2.4E+02 | 2.4E+02 | 6.3E-01 | 4.7E-01 | 8.0E-01 | 7.1E-01 | 6.8E-01 | 7.3E-01 | 7.1E-01 | 7.5E-01 |
| 100 | 500 | 1.3E+02 | 1.3E+02 | 7.6E-01 | 6.6E-01 | 6.0E-01 | 1.1E+00 | 7.6E-01 | 9.2E-01 | 6.7E-01 | 6.6E-01 |
| 100 | 1000 | 2.0E+02 | 2.0E+02 | 6.6E-01 | 6.8E-01 | 1.3E+00 | 6.8E-01 | 6.7E-01 | 1.2E+00 | 5.7E-01 | 6.0E-01 |
| 200 | 100 | 1.3E+02 | 1.3E+02 | 6.0E-01 | 5.9E-01 | 9.2E-01 | 6.9E-01 | 6.9E-01 | 9.3E-01 | 4.9E-01 | 5.3E-01 |
| 200 | 200 | 2.3E+02 | 2.3E+02 | 4.9E-01 | 7.6E-01 | 8.0E-01 | 4.6E-01 | 5.3E-01 | 7.8E-01 | 5.9E-01 | 5.1E-01 |
| 200 | 500 | 2.3E+02 | 2.3E+02 | 5.9E-01 | 4.9E-01 | 8.3E-01 | 6.1E-01 | 5.2E-01 | 8.5E-01 | 5.5E-01 | 4.5E-01 |
| 200 | 1000 | 1.4E+02 | 1.4E+02 | 7.1E-01 | 7.5E-01 | 8.7E-01 | 6.4E-01 | 6.7E-01 | 8.2E-01 | 5.5E-01 | 5.2E-01 |
| 500 | 100 | 2.3E+02 | 2.3E+02 | 6.6E-01 | 6.1E-01 | 7.5E-01 | 8.0E-01 | 6.1E-01 | 6.6E-01 | 5.1E-01 | 3.6E-01 |
| 500 | 200 | 2.2E+02 | 2.2E+02 | 6.1E-01 | 7.8E-01 | 7.7E-01 | 4.1E-01 | 7.6E-01 | 6.1E-01 | 4.0E-01 | 3.4E-01 |
| 500 | 500 | 1.8E+02 | 1.8E+02 | 7.8E-01 | 5.3E-01 | 7.0E-01 | 6.6E-01 | 4.6E-01 | 1.1E+00 | 4.4E-01 | 4.3E-01 |
| 500 | 1000 | 2.2E+02 | 2.2E+02 | 6.2E-01 | 8.5E-01 | 9.1E-01 | 6.6E-01 | 8.9E-01 | 7.8E-01 | 4.2E-01 | 4.2E-01 |
| 1000 | 100 | 1.7E+02 | 1.7E+02 | 5.2E-01 | 6.3E-01 | 6.1E-01 | 7.1E-01 | 7.8E-01 | 7.4E-01 | 6.0E-01 | 4.2E-01 |
| 1000 | 200 | 1.3E+02 | 1.3E+02 | 7.2E-01 | 5.9E-01 | 7.1E-01 | 6.4E-01 | 6.2E-01 | 6.5E-01 | 4.8E-01 | 4.7E-01 |
| 1000 | 500 | 1.5E+02 | 1.5E+02 | 7.5E-01 | 5.1E-01 | 5.9E-01 | 5.1E-01 | 4.9E-01 | 5.8E-01 | 3.1E-01 | 2.9E-01 |
| 1000 | 1000 | 2.3E+02 | 2.3E+02 | 7.2E-01 | 6.5E-01 | 7.2E-01 | 6.9E-01 | 5.5E-01 | 7.5E-01 | 3.6E-01 | 3.7E-01 |

Table 7: Relative efficiency in S2 (for first half of the parameter elements), as measured by the ratio of the ML MSE to the MPL MSE for each parameter element, computed over $N = 100$ repetitions. Here, $aEb = a \times 10^b$.

| $m$ | $n$ | $\pi_1$ | $\pi_2$ | $\pi_3$ | $\pi_4$ | $\beta_{10}$ | $\beta_{11}$ | $\beta_{12}$ | $\beta_{20}$ | $\beta_{21}$ | $\beta_{22}$ |
|---|---|---|---|---|---|---|---|---|---|---|---|
| 100 | 100 | 3.1E+02 | 3.5E+02 | 1.0E+03 | 6.9E+02 | 8.1E-01 | 2.0E+00 | 1.6E+00 | 1.1E+00 | 1.1E+00 | 9.1E-01 |
| 100 | 200 | 1.6E+02 | 2.0E+02 | 7.4E+02 | 8.9E+02 | 1.2E+00 | 1.7E+00 | 1.2E+00 | 1.2E+00 | 1.1E+00 | 6.7E-01 |
| 100 | 500 | 8.7E+01 | 8.8E+01 | 6.8E+02 | 7.2E+02 | 9.0E-01 | 1.8E+00 | 9.6E-01 | 1.0E+00 | 1.9E+00 | 7.6E-01 |
| 100 | 1000 | 3.6E+01 | 5.2E+01 | 8.2E+02 | 9.9E+02 | 8.2E-01 | 2.0E+00 | 7.5E-01 | 6.9E-01 | 1.2E+00 | 8.5E-01 |
| 200 | 100 | 2.9E+02 | 2.3E+02 | 9.4E+02 | 7.8E+02 | 7.8E-01 | 1.5E+00 | 6.7E-01 | 7.4E-01 | 1.0E+00 | 5.3E-01 |
| 200 | 200 | 1.5E+02 | 1.9E+02 | 1.1E+03 | 1.1E+03 | 6.6E-01 | 1.8E+00 | 5.1E-01 | 6.4E-01 | 1.1E+00 | 4.5E-01 |
| 200 | 500 | 6.2E+01 | 8.2E+01 | 6.2E+02 | 7.4E+02 | 7.8E-01 | 1.5E+00 | 6.7E-01 | 6.4E-01 | 1.3E+00 | 5.5E-01 |
| 200 | 1000 | 3.2E+01 | 4.2E+01 | 7.6E+02 | 8.4E+02 | 6.8E-01 | 1.1E+00 | 5.1E-01 | 5.9E-01 | 7.0E-01 | 5.2E-01 |
| 500 | 100 | 2.8E+02 | 3.5E+02 | 8.5E+02 | 1.2E+03 | 7.4E-01 | 1.4E+00 | 6.6E-01 | 8.9E-01 | 8.3E-01 | 4.3E-01 |
| 500 | 200 | 1.3E+02 | 1.7E+02 | 7.8E+02 | 7.9E+02 | 5.7E-01 | 1.1E+00 | 4.1E-01 | 5.0E-01 | 7.7E-01 | 4.2E-01 |
| 500 | 500 | 7.8E+01 | 7.2E+01 | 1.2E+03 | 1.6E+03 | 9.0E-01 | 1.1E+00 | 5.2E-01 | 7.8E-01 | 9.6E-01 | 4.0E-01 |
| 500 | 1000 | 3.4E+01 | 4.3E+01 | 1.2E+03 | 1.0E+03 | 9.8E-01 | 1.0E+00 | 4.1E-01 | 9.9E-01 | 6.9E-01 | 3.8E-01 |
| 1000 | 100 | 3.8E+02 | 2.8E+02 | 1.4E+03 | 1.4E+03 | 6.0E-01 | 7.7E-01 | 2.9E-01 | 6.0E-01 | 7.5E-01 | 3.3E-01 |
| 1000 | 200 | 1.5E+02 | 1.5E+02 | 1.4E+03 | 1.0E+03 | 6.8E-01 | 1.2E+00 | 3.8E-01 | 8.7E-01 | 6.3E-01 | 2.8E-01 |
| 1000 | 500 | 7.1E+01 | 1.1E+02 | 1.2E+03 | 1.5E+03 | 8.1E-01 | 9.3E-01 | 2.3E-01 | 7.2E-01 | 6.4E-01 | 2.0E-01 |
| 1000 | 1000 | 3.6E+01 | 3.6E+01 | 8.7E+02 | 1.1E+03 | 8.5E-01 | 1.2E+00 | 2.6E-01 | 8.5E-01 | 6.8E-01 | 2.6E-01 |

Table 8: Relative efficiency in S2 (for second half of the parameter elements), as measured by the ratio of the ML MSE to the MPL MSE for each parameter element, computed over $N = 100$ repetitions. Here, $aEb = a \times 10^b$.

| $m$ | $n$ | $\beta_{30}$ | $\beta_{31}$ | $\beta_{32}$ | $\beta_{40}$ | $\beta_{41}$ | $\beta_{42}$ | $\sigma_1^2$ | $\sigma_2^2$ | $\sigma_3^2$ | $\sigma_4^2$ |
|---|---|---|---|---|---|---|---|---|---|---|---|
| 100 | 100 | 7.9E-01 | 6.9E-01 | 1.3E+00 | 1.2E+00 | 9.0E-01 | 2.1E+00 | 6.5E-01 | 6.0E-01 | 6.0E-01 | 6.7E-01 |
| 100 | 200 | 8.4E-01 | 8.9E-01 | 2.0E+00 | 7.9E-01 | 5.0E-01 | 1.1E+00 | 7.3E-01 | 7.2E-01 | 6.2E-01 | 6.9E-01 |
| 100 | 500 | 9.7E-01 | 6.3E-01 | 1.4E+00 | 9.7E-01 | 7.9E-01 | 1.7E+00 | 4.8E-01 | 4.3E-01 | 6.1E-01 | 8.2E-01 |
| 100 | 1000 | 1.5E+00 | 5.8E-01 | 1.2E+00 | 1.1E+00 | 7.9E-01 | 1.3E+00 | 4.8E-01 | 3.4E-01 | 8.1E-01 | 7.0E-01 |
| 200 | 100 | 9.0E-01 | 7.5E-01 | 1.5E+00 | 6.6E-01 | 7.9E-01 | 1.5E+00 | 6.7E-01 | 4.7E-01 | 5.6E-01 | 6.9E-01 |
| 200 | 200 | 6.9E-01 | 5.7E-01 | 1.8E+00 | 5.1E-01 | 6.8E-01 | 9.2E-01 | 5.1E-01 | 5.1E-01 | 5.8E-01 | 5.3E-01 |
| 200 | 500 | 5.4E-01 | 4.1E-01 | 1.2E+00 | 7.9E-01 | 3.6E-01 | 1.1E+00 | 3.8E-01 | 3.4E-01 | 6.6E-01 | 7.2E-01 |
| 200 | 1000 | 8.4E-01 | 3.0E-01 | 8.4E-01 | 7.7E-01 | 3.0E-01 | 6.9E-01 | 2.7E-01 | 1.9E-01 | 5.2E-01 | 6.1E-01 |
| 500 | 100 | 9.8E-01 | 4.9E-01 | 8.5E-01 | 8.6E-01 | 3.2E-01 | 6.9E-01 | 4.8E-01 | 4.4E-01 | 4.7E-01 | 4.4E-01 |
| 500 | 200 | 5.8E-01 | 2.9E-01 | 6.7E-01 | 9.0E-01 | 2.6E-01 | 7.6E-01 | 3.5E-01 | 2.7E-01 | 7.4E-01 | 6.4E-01 |
| 500 | 500 | 8.4E-01 | 1.8E-01 | 7.3E-01 | 9.1E-01 | 2.2E-01 | 7.5E-01 | 2.4E-01 | 1.7E-01 | 5.8E-01 | 3.8E-01 |
| 500 | 1000 | 5.4E-01 | 2.2E-01 | 2.9E-01 | 7.9E-01 | 2.3E-01 | 5.6E-01 | 1.5E-01 | 1.1E-01 | 5.5E-01 | 4.5E-01 |
| 1000 | 100 | 9.4E-01 | 4.4E-01 | 8.0E-01 | 1.0E+00 | 3.6E-01 | 7.6E-01 | 3.6E-01 | 3.1E-01 | 1.0E+00 | 1.1E+00 |
| 1000 | 200 | 8.2E-01 | 1.8E-01 | 5.5E-01 | 7.4E-01 | 2.7E-01 | 4.0E-01 | 3.9E-01 | 1.8E-01 | 6.5E-01 | 5.2E-01 |
| 1000 | 500 | 8.4E-01 | 1.5E-01 | 3.7E-01 | 8.4E-01 | 1.8E-01 | 4.0E-01 | 1.7E-01 | 9.2E-02 | 5.9E-01 | 5.7E-01 |
| 1000 | 1000 | 9.8E-01 | 8.6E-02 | 2.9E-01 | 9.9E-01 | 1.2E-01 | 3.1E-01 | 6.0E-02 | 7.4E-02 | 4.6E-01 | 4.8E-01 |

the concordance is perfect for $m = 500, 1000$ and in S2, the concordance is perfect for $m = 1000$, across all values of $n$. This is a good result since the pseudoallocation was considered for use in large $m$ scenarios.

Lastly, it follows from the general theory of PL estimation that there is an efficiency loss due to using MPL estimation, as compared to ML estimation (cf. Cox & Reid (2004) and Kenne Pagui et al. (2015)). The results from Tables 6–8 are in accordance with the general theory, as the large majority of MSE ratios are less than 1. However, we note that the MSE ratios of the mixing proportions are all greater than 1. The apparent super-efficiency of the MPL estimates of the mixing proportions may be due to the fact that one could interpret each individual PL function as an approximate joint density of $m - p$ short time series that arise from $g$-component mixture models with common mixing proportions.

# 6 Example Application

To demonstrate the application of our methodology, we consider an analysis of a time series dataset arising from the fMRI of an individual in the resting-state. The dataset was obtained as part of the event-related task-based study in Orban et al. (2015).

## 6.1 Data Description

In this analysis, we use the resting-state fMRI time series of a single subject (26 years-old male), taken from an fMRI study (Orban et al., 2015). The data were acquired with consent from the individual, after approval by the ethics committee at the Research Center of the Geriatric Institute, University of Montreal, Canada. The subject was right-handed and had no history of neurological or psychological disorders.

The brain imaging data were acquired on a 3-T MRI scanner (Magnetom Tim Trio, Siemens) with a 12-channel head coil. The image used in this experiment has spatial resolution $53 \times 64 \times 46$ voxels ($n = 56470$ voxels after inclusive masking gray-matter brain voxels; individual voxels have volume $3 \times 3 \times 4$ millimeters cubed), and a temporal resolution of $m = 300$ volumes (repetition time of 2000 milliseconds). Data were preprocessed with the NIAK software (http://simexp.github.io/niak/); see also Bellec et al. (2012). The time series at each voxel, $\boldsymbol{y}_s$ for $s = 1, ..., n$, is mean normalized and detrended (i.e. each time series consists of the residuals of an ordinary least-square regression).

## 6.2 MoAR Estimation

Following the analysis in Orban et al. (2015), we fit an MoAR $(4, 10)$ model to the data. Here, we note that $g = 4$ corresponds with the number of clusters reported in Orban et al. (2015, Fig. 1), and we found that $p = 10$ was sufficiently rich for the modeling of fMRI time series. The estimated parameter vectors are provided in Table 9. We have ordered the class labels with respect to the size of the component probability estimates $\hat{\pi}_i$.

Table 9: Parameter estimates of an MoAR $(4, 10)$ model of the time series arising from a resting-state fMRI.

| $i =$ | 1 | 2 | 3 | 4 |
|---|---|---|---|---|
| $\hat{\beta}_{i0}$ | 0.044 | 0.023 | -0.019 | -0.024 |
| $\hat{\beta}_{i1}$ | 0.097 | 1.077 | 0.651 | 0.329 |
| $\hat{\beta}_{i2}$ | 0.141 | -0.028 | 0.014 | -0.010 |
| $\hat{\beta}_{i3}$ | 0.008 | -0.390 | -0.118 | 0.136 |
| $\hat{\beta}_{i4}$ | -0.133 | 0.254 | -0.010 | -0.183 |
| $\hat{\beta}_{i5}$ | -0.003 | -0.120 | -0.022 | 0.135 |
| $\hat{\beta}_{i6}$ | 0.033 | 0.047 | 0.008 | -0.056 |
| $\hat{\beta}_{i7}$ | -0.066 | -0.180 | -0.097 | 0.083 |
| $\hat{\beta}_{i8}$ | -0.015 | 0.175 | 0.0220 | -0.033 |
| $\hat{\beta}_{i9}$ | 0.021 | -0.165 | -0.041 | 0.076 |
| $\hat{\beta}_{i10}$ | -0.021 | -0.015 | -0.031 | 0.048 |
| $i =$ | 1 | 2 | 3 | 4 |
| $\hat{\sigma}_i^2$ | 56.190 | 9.845 | 6.600 | 6.0657 |
| $i =$ | 1 | 2 | 3 | 4 |
| $\hat{\pi}_i$ | 0.136 | 0.262 | 0.264 | 0.338 |

### 6.3 Clustering of Voxels

Using the parameter estimates from Table 9, we cluster the voxels into the $g = 4$ classes. We visualize the clustering at the mid-coronal, mid-horizontal, and mid-sagittal slices, as well as provide the variance-over-time of the voxel intensities (i.e. variance of the time series at each voxel) for the respective slices, for reference, in Figure 2. A point-wise mean and 95% confidence interval of the time series from each of the clusters and 200 voxels that are allocated to each cluster are graphed in Figure 3.

## 6.4 Discussion

We find it encouraging to observe that the clustering is overall symmetric between with respect to the left and right brain hemispheres, as can be observed from an inspection of subplots A1 and A2 from Figure 2. Furthermore, even without smoothing, the clusters across A1–A3 appear to be contiguous, which indicates that adjacent regions of the brain behave similarly at rest, as would be anticipated given the higher strength of homotopic functional brain connections. Furthermore, we see that the majority of the highest variance regions (as observable in subplots B1–B3) appear to be allocated to cluster 4. Thus, the MoAR clustering agrees with the sample variance image.

Upon inspecting Figure 3, the behaviors of the four clusters appear distinct. For example, cluster 3 has a lower variance around the mean, than the other clusters. It will take further scientific investigation to explain the biological relevance of our observations.

We note that although there may be dependence between the image voxels, the conclusion of Theorem 2 still holds under an assumption that the data is strong-mixing instead of IID. A condition that implies strong-mixing is $M$-dependence, whereupon each voxel depends on only a finite number of other voxels within the image [cf. Bradley (2005)].

If one wishes to explicitly account for the dependence between voxels,

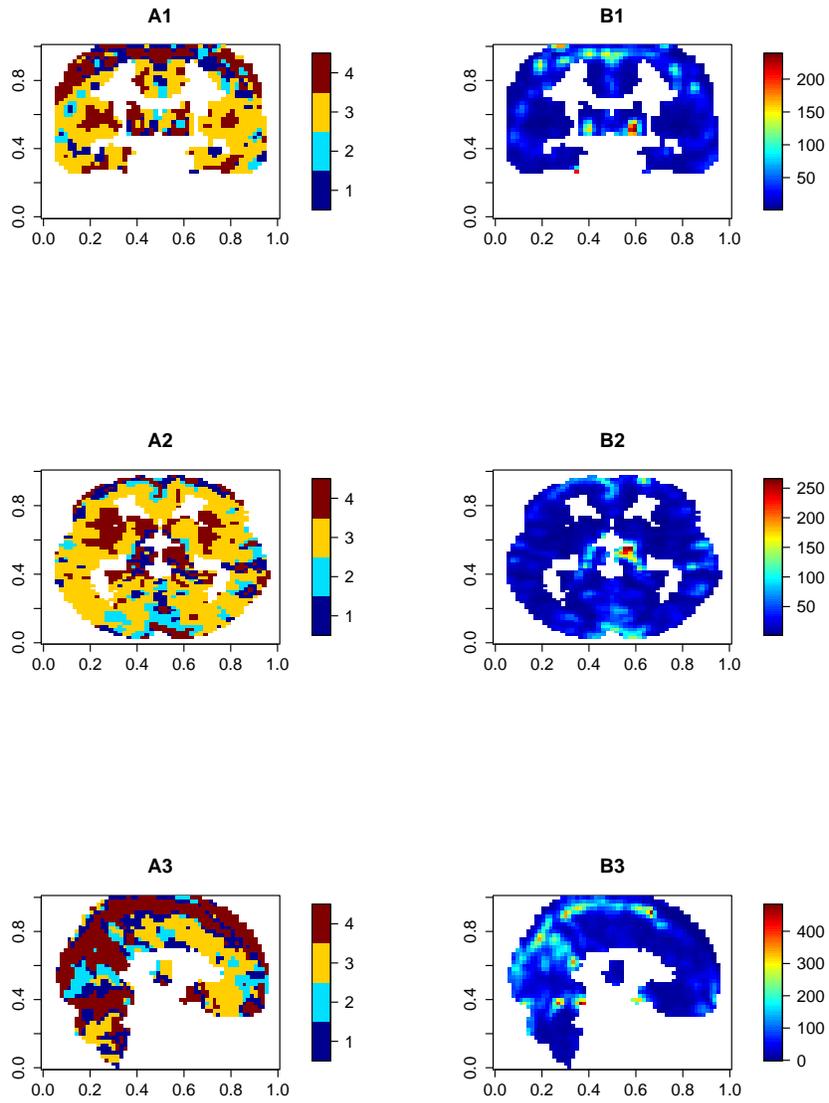

Figure 2: A1–A3 are the visualizations of the clustering at the mid-coronal, mid-horizontal, and mid-sagittal slices, respectively. B1–B3 are the visualizations of the variance image at the respective slices to A1–A3.

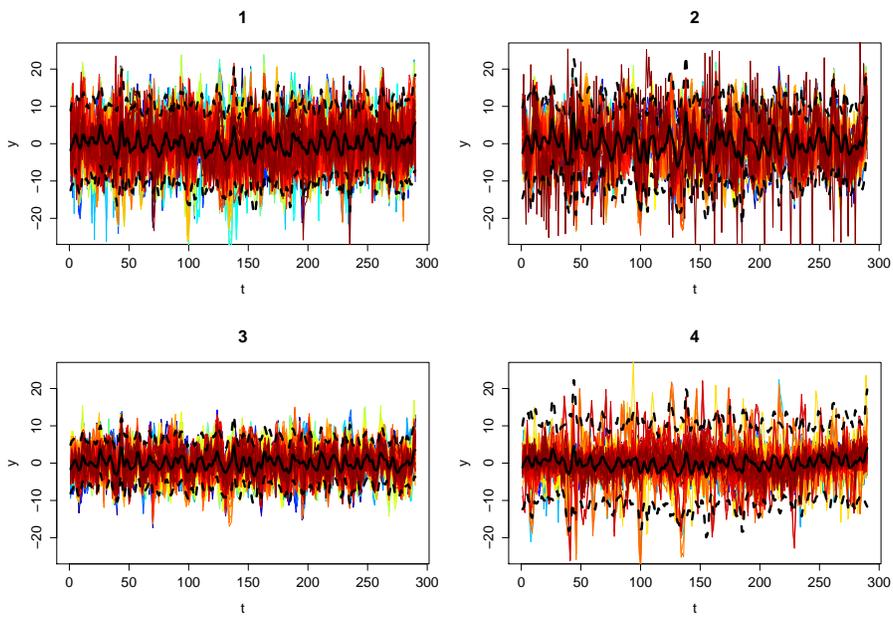

Figure 3: Mean and 95% confidence intervals for each of the four clusters from Section 6.2. The solid line indicates the point-wise mean and the dashed line indicates the 95% point-wise confidence interval, in each plot. The 200 time series belonging to each cluster are plotted as colored lines.

then the Markov random field (MRF) approach of Nguyen et al. (2016) can be applied to obtain a smoother image. Figure 4 displays slices of an MRF spatially-smoothed version of the clustering from Figure 2. The two clusterings differ at approximately 21% of the voxels and it is debatable as to whether or not spatial smoothing is necessary.

# 7 Conclusions

In this article, we discussed the numerical problem inherent in the evaluation of expressions of the form (1), which arise in the ML estimation of MoAR models. In order to circumvent this problem we considered instead the MPL estimator.

An EM algorithm was constructed for the computation of the MPL estimate. It was established that this algorithm increases the PL function after each iteration and the sequence of iterates so produced converges to a stationary point of the log PL function. Furthermore, the MPL estimator was shown to be consistent.

Model-based clustering via the MoAR model requires the evaluation of estimated *a posteriori* probability terms that require the computation of expressions of form (1). To circumvent the evaluation of such expressions, we propose a pseudoallocation rule as an approximation to the usual plugin version of the Bayes' rule.

To assess the performance of the MPL estimator, we performed a number of simulation studies. We found that the MPL estimates converged in MSE to the true parameter, as $n$ increases, as established via the consistency result. However, like other PL estimators, the MPL suffers in efficiency when compared to the ML estimator for the same problem. Surprisingly, we found that the MPL estimates of the mixing proportions $\pi_i$ always had smaller mean squared errors, which is an interesting result that warrants future study.

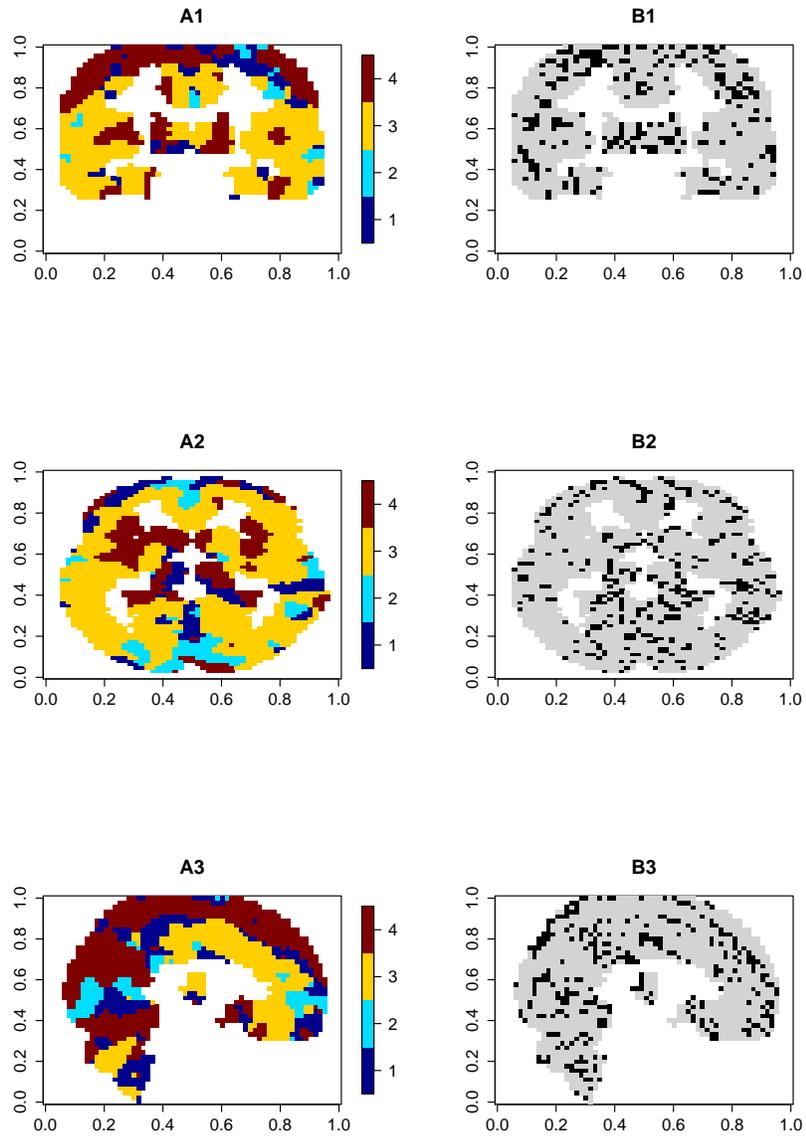

Figure 4: A1–A3 are the visualizations of the spatially-smoothed clustering at the mid-coronal, mid-horizontal, and mid-sagittal slices, respectively. B1–B3 are the visualizations the locations where the smoothed and original clustering differ the respective slices to A1–A3. Here, black indicates a difference.

In addition to our study of the parameter estimates, we also found that the pseudoallocation rule increased in concordance with the Bayes' rule as $m$ increased. This is a useful result as the MPL estimator becomes more useful as the length of the time series increases.

Finally, we demonstrated the methodology developed in this article in an analysis of resting-state fMRI time series of a single individual. The MoAR-based clustering yielded results that are both biologically plausible and were in agreement with the variance-over-time at each voxel.

## Acknowledgments

HDN thanks the CRIUGM at the University of Montreal for their hospitality during the process of writing this article. HDN, GJM, and ALJ thank the Australian Research Council for their financial support.